\documentstyle[aps]{revtex} 

\input epsf
\begin{document}
\twocolumn[
\title{Probing Polarized Parton Distributions
with Meson Photoproduction}
\author{
Andrei Afanasev{\footnotemark}
}
\address{
Thomas Jefferson National Accelerator Facility,
12000 Jefferson Avenue, Newport News, VA 23606\\
and Department of Physics, Hampton University, Hampton,
VA 23668}
\author{
Carl E. Carlson and Christian Wahlquist
}
\address{
Physics Department, College of William and Mary,
Williamsburg, VA 23187
}
\date{December 30, 1996}
\maketitle
\parshape=1 0.75in 5.5in \indent
{\small Polarization asymmetries in photoproduction of
high transverse momentum mesons are a flavor sensitive
way to measure the polarized quark distributions.  We
calculate the expected asymmetries in several models,
and find that the asymmetries are significant and also
significantly different from model to model.  Suitable
data may come as a by-product of deep inelastic
experiments to measure $g_1$ or from dedicated
experiments.       \vglue -20pt}

\widetext \vglue -6.2cm  \hfill JLAB-THY-96-18
\vglue 5.7cm \narrowtext
]


\footnotetext{\vglue -25pt $^*$On leave from Kharkov
Institute  of Physics and Technology, Kharkov, Ukraine.}

\section{Motivation}

In this note, we will describe a flavor sensitive tool
for measuring polarized quark distributions, namely
photoproduction off nucleons of mesons with high
transverse momentum, using polarized initial states.

Presently, information on polarized quark distributions
comes from deep inelastic electron or muon scattering
with polarized beams and targets~\cite{g1}.  Single arm
measurements of $g_1$ give information about a
linear combination of polarized quark distributions.
Obtaining polarized distributions of individual flavors
from this data requires extra theoretical input in the
analysis.  Recently, coincidence measurements of
$\vec \ell\ \vec p (\vec d) \rightarrow \ell\, \pi^\pm X$
have been reported~\cite{adeva}. This data, for a proton
or deuteron target, gives different linear combinations
of up and down quark polarized distributions, allowing a
flavor decomposition without further theoretical
input~\cite{flavor}.

The process we will discuss,
$\vec \gamma\, \vec p \rightarrow M\, X$ (where the
photon is real, targets other than protons are possible,
and $M$ is a meson), gives a complementary way to find the
polarized quark distributions.  The perturbative QCD that
we use in the analysis is justified on the basis of high
meson transverse momentum, rather than by high virtuality
of an exchanged photon,  and the experiment is a single arm
experiment rather than a coincidence one.  Good data can
in fact come as a by-product of a $g_1$ experiment since
the detectors that measure the final electron or muon can
also pick up charged hadrons;  recall that if the final
lepton is not measured, the form of the cross section
ensures that the virtuality of the exchanged photon will
be rather low on the average.

\begin{figure}

\vglue -0.1in
\hskip 0.9 in \epsfysize 1.1 in \epsfbox{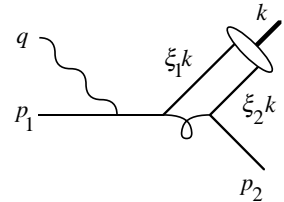}

\caption{One lowest order perturbative diagram for
direct photoproduction of mesons from a quark.  There
are four diagrams total, corresponding to the four places a
photon may be attached to a quark line.}

\label{subproc}
\end{figure}

At high enough transverse momentum, mesons are directly
produced by short range processes illustrated in
Fig.~\ref{subproc}.  These processes are amenable to
perturbative QCD calculation~\cite{bgh,cw93} and produce
mesons that are kinematically isolated in the direction
they emerge.  The direct processes possess several
important features.

One important feature is that the momentum fraction of the
active quark is immediately obtainable from experimentally
measurable quantities.  This is like the situation in
deep inelastic lepton scattering, where experimenters
can measure $Q^2$ and $\nu$ and determine the quark
momentum fraction by $x = Q^2/2m_N \nu$.  In the present
case, we define the Mandelstam variables using
$p$, $q$, and $k$, the momenta of the proton (or other
target hadron), the photon, and the meson,
respectively,
\begin{eqnarray}
s &=& (p+q)^2 , \nonumber \\
t &=& (q-k)^2 , \nonumber \\
u &=& (p-k)^2 .
\end{eqnarray}

\noindent These are all experimentally measurable
quantities.  We can show that, neglecting masses,
\begin{equation}
x = {-t \over s+u}.
\label{eks}
\end{equation}

The second important feature is that the polarization
asymmetry of meson production is known via easy
calculation at the subprocess level.  For example, if
$R$ and $L$ represent photon helicities and $\pm$
represent quark helicities, then one polarization
asymmetry is
\begin{equation}
\widehat{E} \equiv {{{d\hat{\sigma}_{R+}\over dt} -
         {d\hat{\sigma}_{R-}\over dt}}
   \over
   {{d\hat{\sigma}_{R+}\over dt} +
         {d\hat{\sigma}_{R-}\over dt}}}
  = {{\hat{s}^2 - \hat{u}^2} \over
           {\hat{s}^2 + \hat{u}^2}},
\label{asym}
\end{equation}

\noindent where the carets represent subprocess
quantities. (We should emphasize again that internal
quantities such as $\hat{s}$ or $\hat{u}$ are all
determinable from experimentally measurable quantities
if the direct meson production is dominant.)  The
measured asymmetry then measures what fraction of the
quarks have the same or opposite polarization, for a
measured $x$, as the parent proton.

The third important feature of direct meson production is
that by changing the flavor of the meson we observe we
change the flavor of the quarks that the measurement is
sensitive to.  For example, if we observe a $\pi^+$, only
the $u$ and $\bar d$ quarks will contribute.  If, in
addition, we are in the region where valence quarks dominate,
we can get formulas as simple as
\begin{equation}
\Delta u(x,\mu^2) = u(x,\mu^2) {E \over \widehat{E}},
\label{simple}
\end{equation}

\noindent where $E$ is the measured polarization asymmetry,
$\widehat{E}$ is the calculated polarization asymmetry at the
quark level, $u(x,\mu^2)$ is the by now relatively well
known unpolarized quark distribution for the up quark,
$\mu^2$ is a renormalization scale pertinent for the
process and kinematics at hand, and $\Delta u$ is the
polarized quark distribution that we want to measure.

With sensitivity to jets, one can get similar information
from the fragmentation process, that is, from subprocesses
producing quark and gluon final states with fragmentation
turning the quarks and gluons into jets.  This has the
advantage of having a larger cross section overall.
However, if one has data with a single hadron measured
(perhaps as a by-product of a single arm deep inelastic
lepton scattering experiment), then concentrating on the
region where the direct process dominates will yield the
information about the quark distributions in the target.
And of course, flavor identification is easier for a single
hadron than for a jet.

\section{Calculations}

At the subprocess level the direct production of mesons
proceeds as in Fig.~\ref{subproc}.  For the case that
the incoming photon is circularly polarized and target
quark is longitudinally polarized, one gets to lowest
order
\begin{eqnarray}
{d\hat\sigma(\gamma q \rightarrow M q') \over dt} &=&
  {128 g_F^2 \pi^2 \alpha \alpha_s^2 \over 27 (-t) \hat s^2}
    I_M^2
  \left({e_q \over \hat s} + {e_{q'} \over \hat u} \right)^2
     \nonumber \\[1ex] &\times&
\left[ \hat s^2 + \hat u^2
                 + \lambda h
       \left( \hat s^2 - \hat u^2
       \right)
    \right],
\end{eqnarray}

\noindent where $\lambda$ is the helicity of the photon,
$h$ is twice the helicity of the target quark, and $g_F$ is a
flavor factor from the overlap of the $q \bar q'$ with
the flavor wave function of the meson.  It is unity for
most mesons if the quark flavors are otherwise suitable;
for example
\begin{equation}
g_F = \left\{
\begin{array}{cc}
1/\sqrt{2}  &  {\rm for\ } \pi^0 \\
1          &  {\rm for\ } \pi^+
\end{array}   \right. .
\end{equation}

\noindent The integral $I_M^2$ is given in terms of the
distribution amplitude of the meson
\begin{equation}
I_M = \int {d\xi_1 \over \xi_1} \phi_M(\xi,\mu^2).
\end{equation}

\noindent It is precisely the same integral that appears in
the perturbative calculation of the $\pi^\pm$
electromagnetic form factor or of the $\pi^0 \gamma \gamma$
form factor.  Finally,
\begin{eqnarray}
\hat s = (p_1 + q)^2, \nonumber \\
\hat u = (p_1 - k)^2,
\end{eqnarray}

\noindent and $\hat t$ is the same as $t$.

The helicity dependent asymmetry at the subprocess
level may be immediately read off and was given in
Eqn.~(\ref{asym}).  The notation ``$E$'' comes from pion
photoproduction work (see for example~\cite{barker});  it is
analogous to $C_{LL}$ or $A_{LL}$ in $pp$ collision studies.

Another possibly useful asymmetry is the single polarization
asymmetry, where the photon has linear polarization
either in or normal to the plane defined by the outgoing
meson.  This asymmetry is (using $\hat\sigma$ for
$d\hat\sigma/dt$)
\begin{equation}
\widehat\Sigma = {\hat\sigma_\| - \hat\sigma_\perp \over
          \hat\sigma_\| + \hat\sigma_\perp}
       = { 2\hat s \hat u \over \hat s^2 + \hat u^2}.
\end{equation}

\noindent It is interesting to note that both $\widehat E$
and $\widehat \Sigma$ are the same as one would obtain for
Compton scattering, $\gamma q \rightarrow \gamma q$, off the
quark.

Within a hadron target, the quark has light cone momentum
fraction $x$, and $p_1 \simeq x p$.  Neglecting masses, one
has
\begin{equation}
\hat s + \hat t + \hat u = 0.
\end{equation}

\noindent Still neglecting masses,  one has that
\begin{equation}
\hat s = x s, \quad
\hat t = t,   \quad
\hat u = x u.
\end{equation}

\noindent Hence the earlier quoted equation~(\ref{eks})
giving $x$ in terms of measurable quantities follows.

To the overall process, the direct subprocess makes a
contribution that requires no integration to evaluate,
\begin{eqnarray}
E_\pi {d\sigma \over d^3k} &=& {s x^2 \over \pi (-t)}
{d\sigma(\gamma p \rightarrow M + X) \over dx\, dt}
\nonumber  \\
  &=& {s x^2 \over \pi (-t)}  \sum_q G_{q/p}(x,\mu^2)
    {d\hat\sigma(\gamma q \rightarrow M q') \over dt},
\end{eqnarray}

\noindent where the helicity summations are tacit.

The helicity dependent asymmetry is reduced from its
subprocess value because not all the quarks in a hadron
have the same polarization as the hadron does.  This is
what allows us to measure the polarized quark
distributions.  Take $\pi^+$ production off a proton target
as an example.  The initial active quark may be either a $u$
or a $\bar d$, and
\begin{eqnarray}
&E(x,u/s)& = \widehat E(u/s)
\nonumber \\ &\times& {
  \left({e_u \over  s} + {e_{d} \over  u} \right)^2
    \Delta u(x)  +
  \left({e_d \over  s} + {e_{u} \over  u} \right)^2
    \Delta \bar d(x)
                \over
  \left({e_u \over  s} + {e_{d} \over  u} \right)^2
     u(x)  +
  \left({e_d \over  s} + {e_{u} \over  u} \right)^2
     \bar d(x)
  }.
\end{eqnarray}

\noindent We have let
\begin{equation}
q(x) = q(x,\mu^2) = G_{q/p}(x,\mu^2)
\end{equation}

\noindent for the unpolarized distributions, and
\begin{equation}
\Delta q(x) = \Delta q(x,\mu^2) =
  G_{q+/p+} - G_{q-/p+}(x,\mu^2).
\end{equation}

\noindent Also, because of the ratio, we can write the
formula in terms of the measured $s$ and $u$ directly
rather than using the subprocess variables.  At higher $x$,
where the valence quarks dominate,  we may drop the $\bar
d$ terms in the above expression and obtain the simple
result~(\ref{simple}) quoted earlier.

\begin{figure}[h]

\hskip 0.3 in  \epsfysize= 2.0 in   \epsfbox{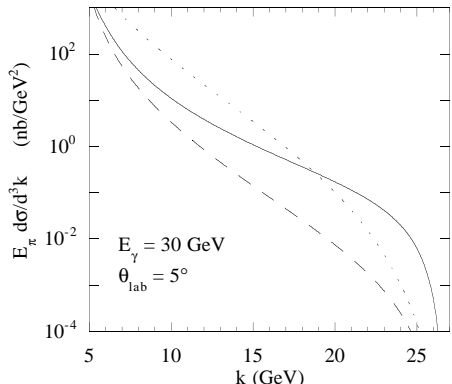}

\caption{Comparing direct and fragmentation photoproduction
of $\pi^+$ off protons, for $E_\gamma = 30$ GeV and
$\theta_{lab} = 5^\circ$.  The solid line is direct
production of $\pi^+$ and the dotted line is fragmentation
production of $\pi^+$.  For reference, the dashed line
shows direct production of $\pi^0$.  The abscissa is
$k = |\vec k|$ in the lab.}

\label{compare}
\end{figure}

We should make a few remarks on the relevance of the direct
process and and its relation to the values of $x$ that are
probed.  The direct process is higher twist, and does not
generally dominate the production of mesons.  It does
dominate in the region of very high transverse momentum.
The main competition is the fragmentation process, where
the observed meson is part of a jet.  The fragmentation
process tails off at the highest transverse momenta simply
because of its implicit requirement that one meson take all
or nearly all the momentum of the quark is unlikely to be
satisfied in a jet.  For illustration, a calculation
comparing direct to fragmentation photoproduction of
$\pi^+$'s off a proton is shown in Fig.~\ref{compare},
based on calculations reported in~\cite{cw93}, and
using the asymptotic distribution amplitude for the
pion (using the Chernyak-Zhitnitsky distribution
amplitude would make the direct process calculation
larger by a factor 25/9).  This particular example is
for photon energy 30 GeV with the pion emerging at
5$^\circ$ in the lab.  The direct
process is larger than the fragmentation process for pion
momenta above 20 GeV;  this corresponds to
$x$ above about 0.24.  Generally, if the meson can be
produced from a valence quark in the target, we will be in
the valence region when the direct process dominates.

\section{Results}

The chief question must be how sensitive a measurement of
$\Delta q$ can be made.  To study this question, we use
three different models or fits to the polarized quark
distributions.  These are the fits of Gehrmann and
Stirling (GS)~\cite{gs96}, of Gl\"uck, Reya, Stratmann, and
Vogelsang (GRSV)~\cite{grsv96}, and a suggestion of Soffer
{\it et al.}~\cite{soffer}.  All fit the available data on
$g_1$ from deep inelastic lepton scattering experiments.
For the first two, we have obtained the renormalization
scale dependent results for the polarized parton
distributions directly from the authors. The
Soffer {\it et al.}\ suggestion relates the polarized and
unpolarized distribution functions, specifically,
\begin{eqnarray}
\Delta u(x) &=& u(x) - d(x) , \nonumber \\
\Delta d(x) &=& - {1\over 3} d(x) ,
\end{eqnarray}

\noindent and other polarized distributions are treated as
small.  When we use the Soffer {\it et al.}\ suggestion,
we team it with the CTEQ~\cite{cteq95} quark
distributions.  In all cases, we set the renormalization
scale $\mu^2$ to
$k_T^2$, where
$k_T$ is the transverse momentum of the produced meson.

\begin{figure}

\hskip 0.2 in  \epsfysize= 2.3 in   \epsfbox{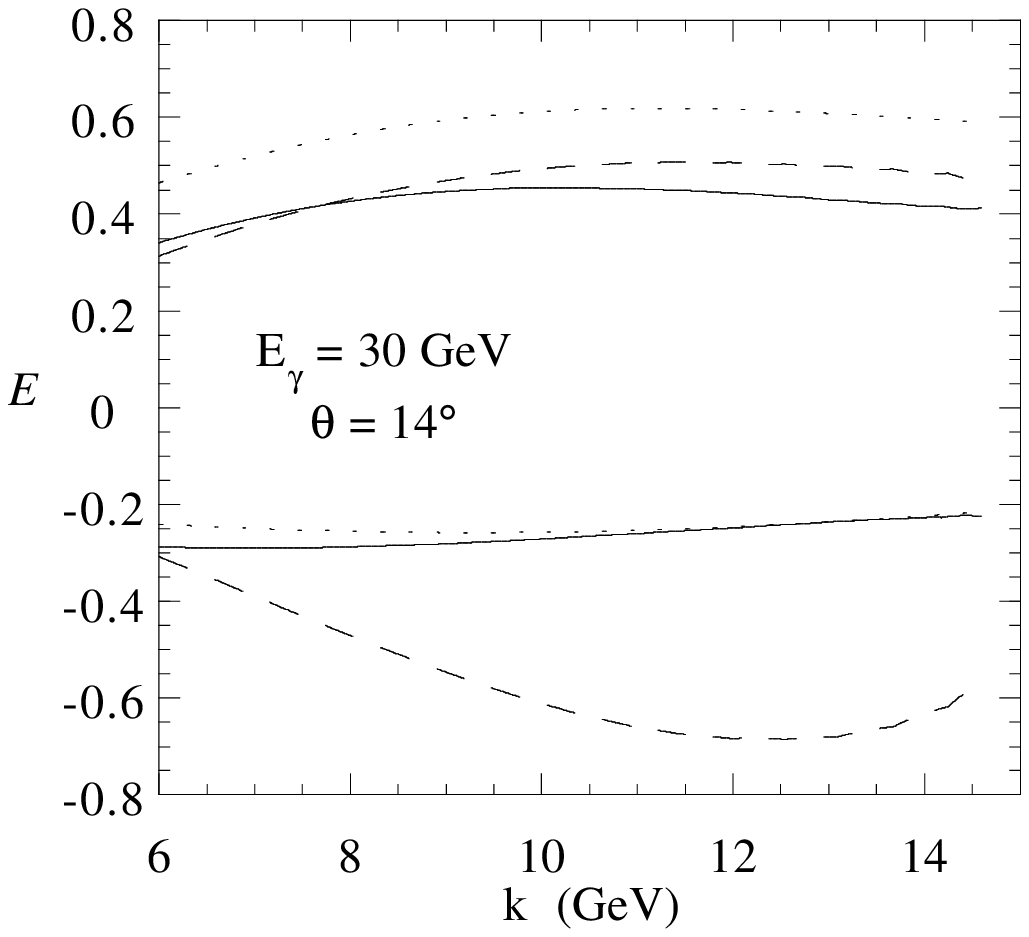}
\vglue -.01in
\hglue .2 in   \epsfysize= 2.3 in    \epsfbox{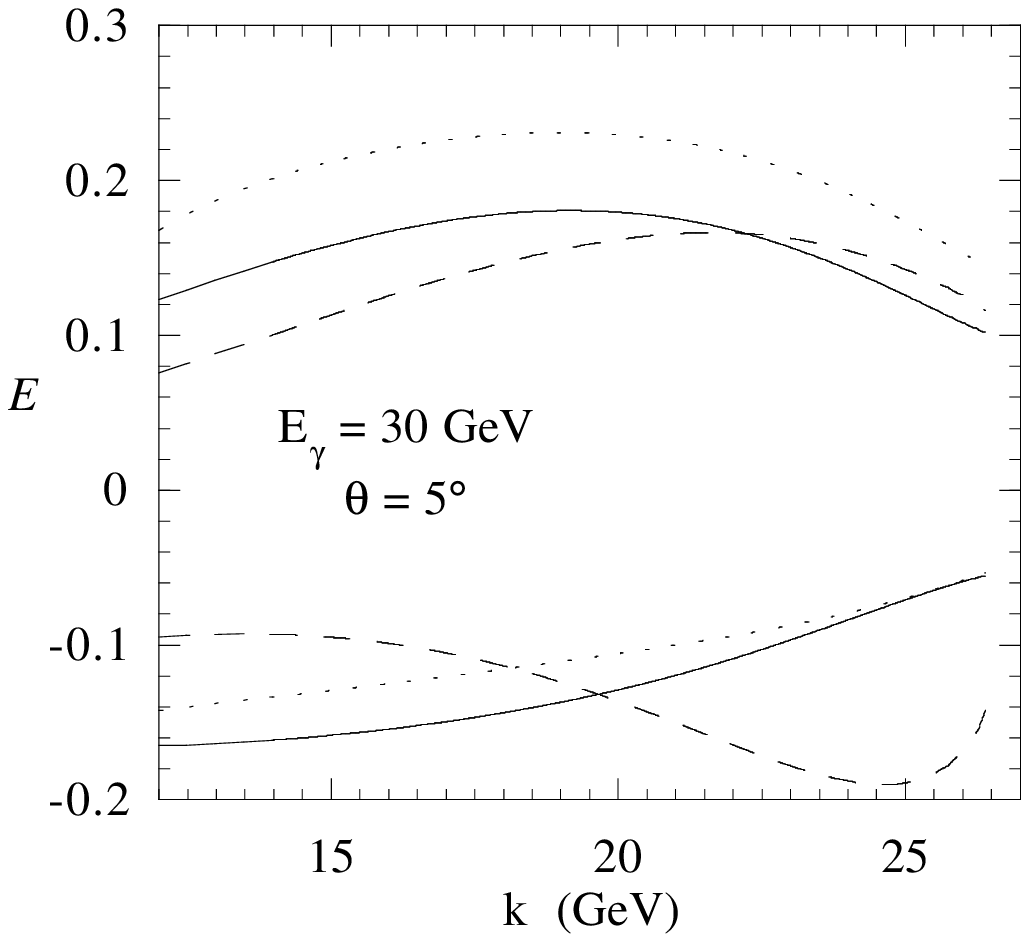}

\caption{The upper three curves are the helicity
dependent asymmetries for $\pi^+$
photoproduction off the proton, for 30 GeV photons and lab
angle 5 or 14$^\circ$.  The solid line is for the GRSV
polarized quark distributions; the dashed line is for
GS; and the dotted line is for the suggestion of
Soffer {\it et al}.  GRSV is about the same as GS for
the $\pi^+$ case.  The lower three curves are for
$\pi^-$ photoproduction. GRSV and GS are well separated in the
$\pi^-$ case.}

\label{asymm}
\end{figure}

\begin{figure}

\hskip 0 in  \epsfysize= 2.3 in   \epsfbox{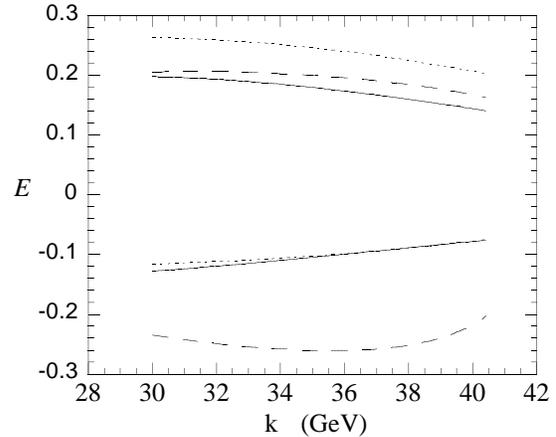}

\caption{Polarization asymmetry results for 50 GeV incoming
electrons, with pions emerging at 5$^\circ$ lab angle and the
electrons not observed.  We have integrated over the
energies and polarizations of the virtual (but on the average
low $Q^2$) photons with weightings as indicated in the
text.  The upper three curves are for the $\pi^+$ and the lower
three curves are for the $\pi^-$.  The solid, dashed, and
dotted lines are for GRSV, GS, and Soffer {\it et al.},
respectively.}

\label{electro}
\end{figure}

The upper part of Fig.~\ref{asymm} shows the asymmetry
$E$ for $\pi^+$ photoproduction off protons.  (Both
Fig.~\ref{asymm} and \ref{electro} are for 100\%
polarization of the beam and target.)  We notice that the
asymmetries are significant, and that the GS and GRSV
polarized distributions give about the same result, but
the Soffer {\it et al.}\ suggestion gives an asymmetry
that is significantly larger.  The other part of
Fig.~\ref{asymm} shows the
$\pi^-$ case.  The asymmetry has the opposite sign because
now the $d$ valence quark dominates, and in all the models
the $u$ is polarized along the direction of the proton
while the $d$ is opposite.  For the $\pi^-$, it is the
GRSV and Soffer {\it et al.}\ results that are about the
same, and the GS is significantly different and usually
larger in magnitude.

Electroproduction with the final electron unobserved is
much akin to photoproduction.  Single arm electroproduction
experiments commonly get such data when a charged hadron
rather than an electron triggers the detector.  The form of
the cross section ensures that low $Q^2$ events dominate if
$Q^2$ is not measured.

Hence electroproduction with just the single hadron
observed usually has the photon nearly on shell, but only
the upper and lower limits of the photon energy are known.
The three models for the polarized quark distributions
still give distinguishable predictions, as we shall show.
For a given electron energy $E_e$, the photon energy
spectrum is given fairly accurately by
\begin{equation}
{dN_\gamma \over dE_\gamma} \propto {1 \over E_\gamma}
\end{equation}

\noindent up to close to the cutoff at $E_\gamma = E_e$.
The polarization of the photon is nearly the polarization
of the electron provided the photon takes most of the
electron's energy.  Polarization details can be found
in~\cite{hm}; most usefully, if $P_\gamma$ and $P_e$ are the
circular and longitudinal polarizations of the photon
and electron, respectively, then
\begin{equation}
{P_\gamma \over P_e} = {y(4-y) \over 4-4y+3y^2}.
\end{equation}

\noindent where $y=E_\gamma/E_e$.

Fig.~\ref{electro} shows polarization asymmetry results
for 50 GeV incoming electrons, with pions emerging at
5$^\circ$ lab angle and photon energies and polarizations
weighted as indicated above. Despite the weighted average
over photon energies, the models still give different
predictions.

Another possibility for producing real photons with
circular or linear polarization is the laser
backscattering technique~\cite{laser}.

\section{Discussion}


\vglue -5pt
A number of questions may be asked about the
applicability of perturbative QCD.  We could of course point
out that we are dependent on ratios of cross sections, so
that many potential problems may cancel out.  However, we
shall address some of the questions more directly.

One simple question is whether the ``$X$'' in
$\gamma + p \rightarrow M + X$ is  out of the resonance
region.  Letting $m_X$ be the mass of the collection of
particles $X$, we should require $m_X > 2$ GeV, and
neglecting the mass of the meson  we have
\begin{equation}
m_X = \sqrt{s+t+u-m_N^2}.
\end{equation}

\noindent For $E_\gamma = 30$ GeV and
$\theta_{lab} = 5^\circ$, the requirement becomes $k < 25$
GeV, which is easy to satisfy.
Lower energies could be more troublesome.  At 12 GeV
incoming photon energy, with a $\pi^+$ exiting at
22$^\circ$, the requirements of dominance of the direct
process and of being out of the resonance region just
leave a window
$4.8 < k < 5.3$ GeV, corresponding to
$0.62 < x < 0.74$.

Another question regards higher twist corrections, for
example, corrections due to the quarks in the pion
having finite momentum transverse to the pion's
overall momentum.  This has been much studied in the
context of the pion electromagnetic form
factor~\cite{listerman}.  As has been remarked, the
integral over the pion's distribution amplitude that
appears here is the same as appears in the pion form
factor.  However~\cite{cw93}, the momentum transfer
scale as judged by how far the transferred gluon is
off shell is much larger in pion photoproduction than
in the pion form factor.  To be more definite, if the
photon attaches to the produced $q \bar q$ pair in
Fig.~(\ref{subproc}) the gluon is off shell on the
average by
\begin{equation}
\langle q_G^2 \rangle = \langle \xi_2 \rangle x u  ,
\end{equation}

\noindent The corresponding quantity if the photon
attaches to the initial quark is larger, though timelike.
The average $\xi_2$ is weighted by the integrand of
$I_\pi$ and is 1/3 and 1/5 for the asymptotic and
Chernyak-Zhitnitsky distribution amplitudes,
respectively.  The pion electromagnetic form factor
involves the distribution amplitude twice, and if the
virtual photon is off shell by an amount $Q^2$, then
the gluon in that process is off shell by
\begin{equation}
\langle q_G^2 \rangle = - \langle \xi_2 \rangle^2 Q^2.
\end{equation}

\noindent Matching the gluon virtualities, there is a
correspondence
\begin{equation}
Q^2 \leftrightarrow {x (-u) \over \langle \xi_2 \rangle}.
\end{equation}

\vglue -5pt
\noindent This means that for the case of incoming 30 GeV
photons and pions out at $5^\circ$, in the region of direct
process dominance and above the resonance region, the
integrals over the distribution amplitude are the same as
those for
$F_\pi(Q^2)$ with
\begin{equation}
15 {\rm\ GeV}^2 < Q^2 < 80 {\rm\ GeV}^2.
\end{equation}

\vglue -5pt
\noindent Thus higher twist effects will be less
significant for measurable photoproduction of high
transverse momentum mesons than for meson form
factors at any currently measured momentum transfers.

Perturbative corrections that are higher order in
$\alpha_s$ have not been calculated.  They may be  calculated
along the lines of~\cite{braaten83} for the
$\pi^0\gamma\gamma$ and or of~\cite{braatentse} for the
$\pi^\pm$ electromagnetic form factors.  For both of these,
using the asymptotic distribution amplitude and a suitable
choice of renormalization scale,  the magnitude of the
correction was about 20\%, decreasing the
$\pi^0\gamma\gamma$ and increasing the $\pi^\pm$
form factors.

One may ask about effects of the
fragmentation process.  Even in the region where the
direct process is dominant, there will be some
contribution to the polarization asymmetry from
fragmentation processes~\cite{peralta}.  We have calculated
corrections to $\Sigma$ and $E$.  For $E$, we find the
corrections are small.  For example, again using $E_\gamma
= 30$ GeV and
$5^\circ$ in the lab outgoing pions, at $k = 22$ GeV
including the fragmentation process increases the
asymmetry by a factor 1.02. (We neglected the
polarization of the gluons; in this region $\gamma g$
fusion is about 30\% of the fragmentation
contribution).  One reason the corrections are small
is that the expected asymmetry for both the
quark-gluon fusion and Compton subprocesses is the
same sign and roughly the same magnitude as for the
direct process.

Regarding the single polarization asymmetry $\Sigma$, for
the direct process, it is always negative and for forward
center of mass angles it is large in magnitude.  As we
move to lower $k$ and the fragmentation process becomes
more important, there is a significant dilution of
$\Sigma$.  The reduction of $\Sigma$ occurs because the
$\gamma g$ fusion process becomes important, and $\Sigma$
for that process is precisely zero.  If $\Sigma$ is
observed and seen to be large, it is one sign that the
direct process is important.  For instance, for the
same $E_\gamma$ and $\theta_{lab}$ as above, the
value of $\Sigma$ at $k = 22$ GeV is $-0.61$,
$-0.94$, and $-0.86$, respectively, for the
fragmentation alone, the direct process alone, and
their properly weighted total (using the Soffer
$et al.$ model).

To conclude, we believe that photoproduction of high
transverse momentum mesons is an accurate and flavor
sensitive way to measure the polarized quark
distributions in the valence region.

\section*{\vglue -25pt acknowledgments}

\vglue -10pt
We thank K. Griffioen and J. Gomez for useful discussions
and T. Gehrmann, M. Stratmann, and J. Qiu for supplying
parton distribution computer codes from the GS, GRSV, and
CTEQ collaborations, respectively. AA thanks the DOE for
support under grant DE--AC05--84ER40150; CEC and CW thank
the NSF for support under grant PHY-9600415.

\vglue -15pt

\end{document}